\documentclass[preprint]{aastex62}

\usepackage{natbib}
\usepackage{graphicx}
\usepackage{epstopdf}
\usepackage{amsmath}

\newcommand{\ISO}{{\it ISO}\ }

\newcommand{\Planck}{{\it Planck}\ }
\newcommand{\Spitzer}{{\it Spitzer}\ }
\newcommand{\WISE}{{\it WISE}\ }
\newcommand{\Gaia}{{\it Gaia}\ }

\newcommand{\um}{$\mu$m}

\shorttitle{B-Field of GF~9-2, L1082C, and GF~9}
\shortauthors{Clemens et al.}

\begin{document}

\title{Magnetic Field Uniformity Across the GF~9-2 YSO, L1082C Dense Core, \\
and GF~9 Filamentary Dark Cloud}

\author[0000-0002-9947-4956]{Dan P. Clemens}
\affiliation{Institute for Astrophysical Research, Boston University,
    725 Commonwealth Ave, Boston, MA 02215}

\author[0000-0002-4958-1382]{A. M. El~Batal}
\affil{Institute for Astrophysical Research, Boston University,
    725 Commonwealth Ave, Boston, MA 02215}


\author[0000-0002-8261-9098]{C. Cerny}
\affil{Institute for Astrophysical Research, Boston University,
    725 Commonwealth Ave, Boston, MA 02215}

\author[0000-0002-3469-5774]{S. Kressy}
\affil{Institute for Astrophysical Research, Boston University,
    725 Commonwealth Ave, Boston, MA 02215}

\author[0000-0001-9915-8147]{G. Schroeder}
\altaffiliation{Boston University Astronomy Department Research Experience for Undergraduates (REU) 2017 summer student}

\affil{Department of Physics and Astronomy,
University of Rochester,
206 Bausch and Lomb Hall,
P.O. Box 270171
Rochester, NY 14627-0171}

\author[0000-0003-2133-4862]{T. Pillai}
\affil{Institute for Astrophysical Research, Boston University,
    725 Commonwealth Ave, Boston, MA 02215}

\correspondingauthor{Dan P. Clemens}
\email{clemens@bu.edu}

\begin{abstract}
The orientation of the magnetic field (B-field) in the filamentary dark cloud GF~9 was traced from
the periphery of the cloud into the L1082C dense core that contains the low-mass, low-luminosity
Class~0 young stellar object (YSO) GF~9-2 (IRAS~20503+6006). 
This was done using SOFIA HAWC+ dust thermal emission polarimetry (TEP) at 
216~\um\ in combination with Mimir near-infrared background starlight polarimetry (BSP) 
conducted at $H$-band (1.6~\um) and $K$-band (2.2~\um). 
These observations 
were augmented with published $I$-band (0.77~\um) BSP and \Planck 850~\um\ TEP
to probe B-field orientations with offset from the YSO in a range
spanning 6000~AU to 3~pc. 
No strong B-field orientation change with 
offset was found, indicating  
remarkable uniformity of the B-field from the cloud edge to the YSO environs. 
This finding disagrees with weak-field models of cloud core and YSO formation. 
The continuity of inferred B-field orientations for
both TEP and BSP probes is strong evidence that both are sampling a common
B-field that uniformly threads the cloud, core, and YSO region. 
Bayesian analysis of 
\Gaia DR2 stars matched to the 
Mimir BSP stars finds a distance to GF~9 of $270 \pm 10$~pc. 
No strong wavelength dependence of B-field orientation angle was found, contrary to 
previous claims.
\end{abstract}


\section{Introduction}

Testing whether magnetic (‘B’) fields are important in star-forming 
processes requires the examination of young systems early in their star-formation histories, 
ideally prior to, or soon after, the onset of outflows and jets, which 
can disturb remnant B-field signatures. The characteristics of the undisturbed B-fields,
especially field orientations and strengths, and the spatial variations of those quantities, are
important tools for testing models of cloud core and embedded star formation.
\added{Many previous studies and surveys have investigated B-fields in star formation
regions, including comparing B-field properties at
small and large size-scales \citep[e.g.][]{Li09,Hull14,Zhang14}, but these have tended to focus
on regions of massive or later-stage star formation.} 
In this study, variations in the plane-of-sky orientation of the B-field were examined
versus offset from a \added{single, isolated, low-mass, young} protostar, across a wide range of 
offsets, to characterize orientation changes associated with cloud-to-core
or core-to-protostar B-field configuration changes early in the formation of a low-mass star.

The laboratory for this study was the sky field containing the ‘Class 0’ young stellar object 
(YSO) IRAS~20503+6006 (aka ``GF~9-2"), which is forming 
in the L1082C (LM~351) dense core in the clumpy filamentary cloud GF~9 \citep{Schneider79}. 
This YSO 
shows a two-component spectral energy distribution \citep{Clemens99}, with a $\sim$50~K 
warm component,  a $\sim$12~K cool dust component, and total luminosity 
of $\sim$0.7~$L_\sun$, for an assumed distance of 440~pc \citep{Dobashi94}. 
A closer distance of 200~pc was favored by \citet{Furuya06}, 
who cited \citet{Wiesemeyer97}, to obtain an even lower YSO luminosity of 0.3~$L_\sun$.

GF~9-2 was first identified as a YSO by \citet{Beichman86}, based on {\it IRAS} fluxes, 
and identified as Class 0 by \citet[][who unsuccessfully searched for CO 
outflows]{Bontemps96}. In the mid-infrared, GF~9-2 was examined using 
the ISOCAM instrument \citep{ISOCAM} aboard the {\it Infrared Space Observatory (ISO)} \citep{ISO}
in mapping by \citet{Wiesemeyer99}, who followed up with 3~mm continuum and 
CS~(2-1) IRAM Plateau de Bure interferometer maps. 

Near-infrared (NIR) imaging \citep{Ciardi98} and CO and CS radio spectral line mapping \citep{Ciardi00} 
were used to study the L1082C dense core in which GF~9-2 is embedded and 
to assess the temperatures, densities, and turbulent states there. These studies found 
T$_{GAS}\sim8$~K, M$_{CORE}\sim50$~M$_\sun$, n$_{H2}\sim$5000~cm$^{-3}$, 
maximum A$_V\sim10$~mag, and $\Delta V\sim0.5-0.9$~km~s$^{-1}$. \citet{Furuya03} 
detected H$_2$O maser emission and argued {\it “GF~9-2 is the lowest luminosity object 
known to possess H$_2$O masers.”} \citet{Furuya06} probed GF~9-2 in radio lines of 
NH$_3$, CCS, HCO+, C$_3$H$_2$, SiO, N$_2$H+, CO (1-0 and 3-2), and in 
3.6~cm, 3~mm, and 350~$\mu$m continuum. No free-free continuum was 
found. Instead, thermal dust emission was found at 3~mm and 350~$\mu$m with weak 
elongations to the southeast and northwest, perpendicular to 
elongations seen in some of the gas tracers. CO outflow was again not found 
and no shock-excited SiO was found in these single-beam observations, 
attesting to the low luminosity and extreme 
youth of the YSO. \citet{Furuya09} found gas infall in HCO+ and 
HCN lines, with velocities of 0.5~km~s$^{-1}$ and an estimated accretion rate of 
$2.5 \times 10^{-5}$~$M_\sun$~yr$^{-1}$. \citet{Furuya14a} concluded, 
from their isothermal 
cylinder model, that the filament out of which GF~9-2 condensed must have 
been supported by turbulent plus magnetic pressures. 
\citet{Furuya14b} observed $^{12}$CO~(3-2) with the submillimeter array (SMA) 
to finally reveal a tiny outflow of size 5~mpc (5 arcsec) with an estimated age
of about 500~years.

These observations show that GF~9-2 and its surrounding dense core are involved in the 
earliest phases of star formation, where gas infall is taking place and central 
luminosity is becoming substantial, and where jets and outflows have just recently
been launched. Characterizing the B-field properties 
of this YSO and its dense core could establish a new paradigm for the conditions
involved in early star formation. 

Dust thermal emission polarimetry (TEP) 
of the GF~9-2/L1082C region was reported by \citet{Clemens99}, based on observations
conducted at 160~\um\ using the ISOPHOT \citep{ISOPHOT} instrument aboard {\it ISO}.
However, the polarization properties found by these observations
have been questioned, based on their poor correspondence to the background 
starlight polarimetry (BSP) position angles (PAs) for stars
observed at 0.77~\um\ in a larger, surrounding field of view, by \citet{Poidevin06} 
and to the BSP of six stars observed by \citet{Jones03} in the $H$-band NIR in the same
field of view as the ISOPHOT observations. 

In particular, the apparent change of polarization 
PA with wavelength, from the optical to NIR to \ISO far-infrared (FIR),
led \citet{Poidevin06} to suggest the presence of a wavelength-dependence. 
Such rotation of linear polarization PA with wavelength is rarely seen, 
but can be a powerful signpost of 
changes in B-field orientations and dust properties along the line-of-sight,
as found by
\citet{Messenger97} in the optical for some Taurus 
dark cloud directions. 
Finding astronomical targets exhibiting PA rotation with wavelength, 
and modeling that behavior, could reveal both B-field and dust 
properties. 
PA rotation occurs because different wavebands have different opacity and 
grain alignment efficiency \citep[e.g.,][]{Lazarian07} functions of 
optical depth, which can manifest in cloud cores
\citep{Whittet08}. Additionally, in the presence of embedded, or nearby external, 
sources of luminosity, anisotropic illumination affects grain alignment 
\citep{Andersson11, Jones15}.   
If the \citet{Poidevin06} finding of PA rotation in L1082C were true, then that dense 
core could be experiencing 
B-field directional changes (for example, trading mostly line-of-sight B-field projection 
for mostly plane-of-sky B-field projection). Alternatively, strong changes in dust grain 
sizes or degrees of alignment with the local B-fields could be taking place. 
A correlation of these possible B-field, or dust, changes with early phases 
of star formation could constitute an important test of dense core growth and/or gas
infall models.

\begin{figure}
\includegraphics{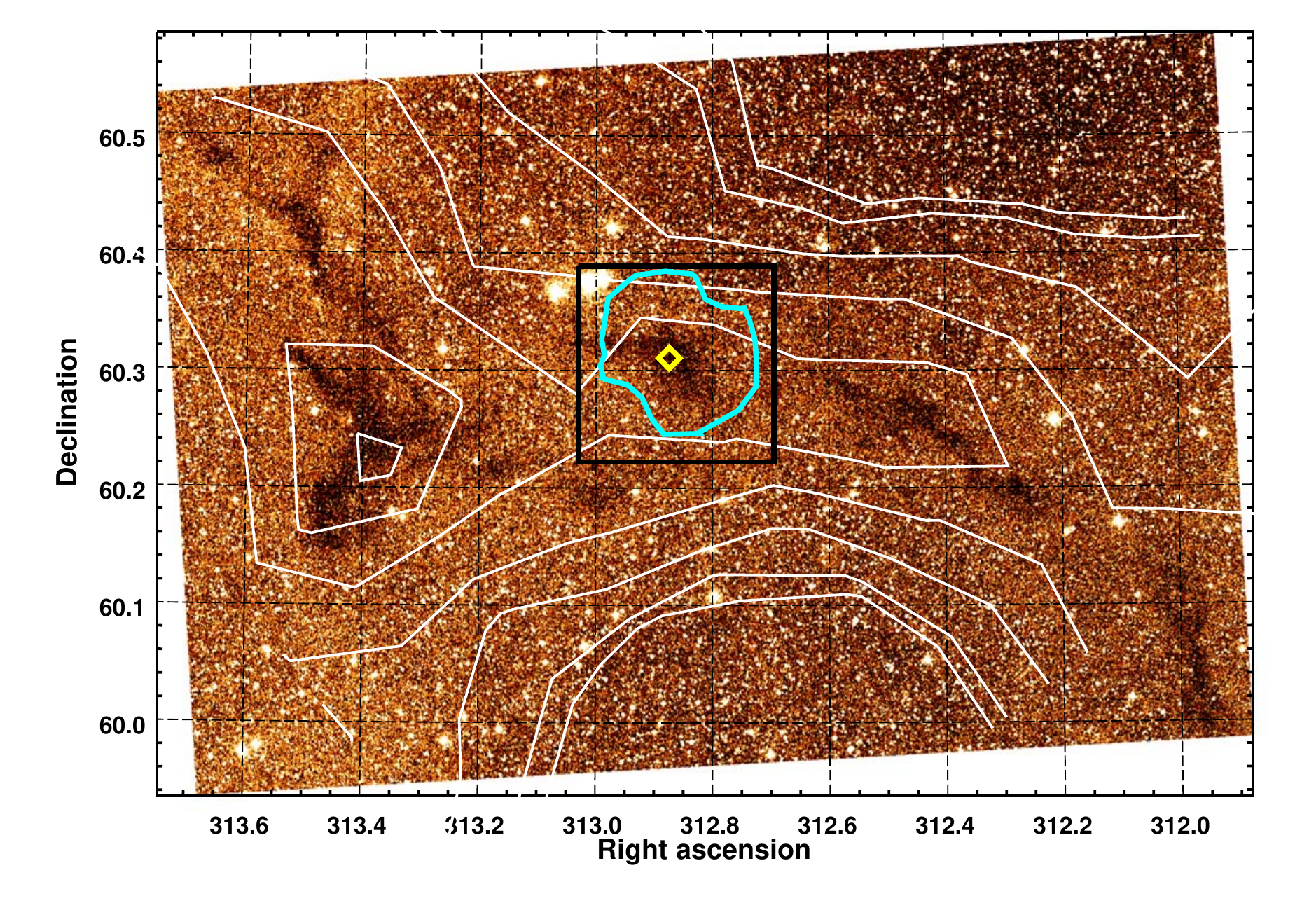}
\caption{\small
Digital Palomar Observatory Sky Survey \citep{DPOSS} representation of GF~9 region. 
White contours show \Planck\ 850~\um\ intensity, cyan polygon represents the 
SOFIA HAWC+ 216~\um\ coverage, the black box shows the Mimir field of view, 
and the yellow diamond shows the location of the
GF~9-2 YSO. The GF~9 filamentary dark cloud
is the faint, mostly starless region that extends from lower right to middle-upper left in the 
image. 
\label{fig_overview}}
\end{figure}

This current study was undertaken to investigate the nature of the B-field properties
near and surrounding the very young protostar GF~9-2 and to test for the presence
of polarization PA rotation with wavelength. To accomplish these two goals, 
observations utilizing the FIR imaging polarimeter
High-resolution Airborne Wideband Camera \citep[HAWC+;][]{HAWC_1, HAWC_2, HAWC+}
aboard the Stratospheric Observatory for Infrared Astronomy \citep[SOFIA;][]{SOFIA_1, SOFIA_2} were combined with NIR imaging polarimetric observations using
the Mimir instrument \citep{Clemens07} of the same field. These observations are
described in the following Section and combined with published optical \citep{Poidevin06}
and \Planck \citep{Planck, Planck_pol} polarization data to examine how B-field orientations change with offset from the YSO and to test for wavelength-dependent polarization
PAs.

\section{Observations}\label{Observations}
The GF~9 filamentary dark cloud is seen in 
Figure~\ref{fig_overview} \citep[also see Figure~1 in][]{Poidevin06} from the Digital Palomar Observatory Sky Survey \citep{DPOSS}
as the band of mostly star-free material that extends from the lower
right to upper left. The locations of the GF~9-2 YSO,
the placement of the $10 \times 10$~arcmin$^2$ field of view (FOV) for the
Mimir instrument, and the extent of the SOFIA HAWC+ coverage are shown as overlays in Figure~\ref{fig_overview}.

\subsection{SOFIA HAWC+ Observations}\label{SOFIA}
SOFIA airborne observations of the GF~9-2 region using HAWC+ were obtained on the nights 
of 2016 December 13 and 14,
at altitudes in excess of 12.5~km, for the SOFIA Cycle~4 program 04\_0026. 
In the $E$-Band (216~\um\ center wavelength,
43~\um\ bandwidth) polarimetry mode of HAWC+, the detector
format consisted of two bolometer arrays of nominal pixel counts $32 \times 40$, and
pixel projected dimensions of 9.33~arcsec per side. Each array was illuminated by an
opposite sense of linear polarization from the polarization beam-splitter, with most
sky directions simultaneously observed by pixels in both arrays. The pixel sampling was 
at half the 
diffraction-limited beamsize (18.2~arcsec FWHM) at this wavelength for the 2.5~m 
clear aperture of the SOFIA telescope. Observations were obtained using a chop-nod
procedure, with identical chop and nod throws of 150~arcsec, tertiary chop frequency of
10~Hz, and telescope nod dwell time of 40~s. A $2 \times 2$ position dither, with motions
equal to three pixels (27~arcsec), was combined with four half-wave-plate (HWP) position 
angles to enable polarimetry. A total integration time of one hour was
achieved for the two observing nights.

The raw data were processed by the HAWC+ instrument team, including corrections
for dead pixels and relative pixel gains as well as instrument and telescope intrinsic polarizations. The
chops, nods, dithers, and HWP orientations were rectified and the resulting images registered
and averaged to produce ``Level 4" (science-quality) data products. These included FITS
images of total intensity (Stokes~$I$), fractional polarization $P$, polarization PA, 
Stokes~$UI$ and Stokes~$QI$, and all uncertainties. Initial examination of these images
revealed high signal-to-noise-ratio (SNR) detection of total intensity in $E$-Band 
across much of the dense cloud core, but no
significantly detected polarization at the native resolution of the detector pixels. The 
successful extraction of polarization from these data is described in 
Section~\ref{Obs_obs}.

\begin{figure}
\includegraphics[trim=-0.75in 0.33in 0 0in, clip, angle=0,scale=0.75]{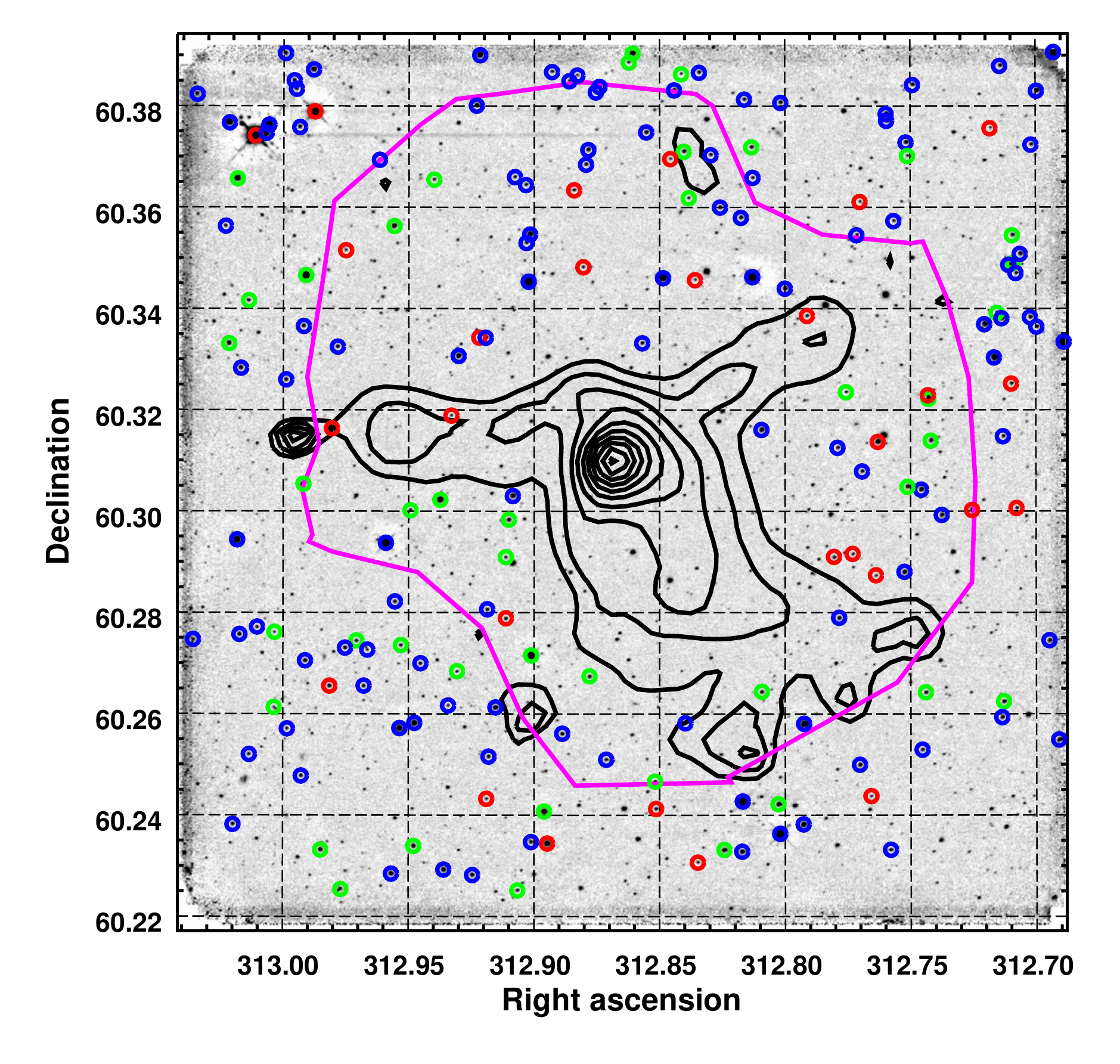}
\caption{\small
Inverted grayscale Mimir image of stacked $K$-band NIR observations of the L1082C core
and GF~9-2 YSO covering the same field of view as delineated by the black square in 
Figure~\ref{fig_overview}. SOFIA HAWC+ total 216~\um\ intensity contours are 
overlaid in black, spanning 20 to 260~MJy~sr$^{-1}$. The extent of the HAWC+ sampled
region is shown as the magenta polygon.
Colored circles identify stars that appear in \Gaia Data Release 2 and have parallax SNR 
greater than two, with colors identifying
distance bins (red for closer than 0.5~kpc,  green for 0.5 to 1.0~kpc, and blue for beyond
1.0~kpc). 
\label{fig_SOFIA_Gaia}}
\end{figure}

Figure~\ref{fig_SOFIA_Gaia} shows black contours of the HAWC+ 216~\um\ total intensity overlaid
on a $10 \times 10$~arcmin$^2$ NIR $K$-band image obtained with Mimir. The outer, magenta
contour identifies the region where the HAWC+ intensity values are valid, 
with reduced SNR near the edges of that region due to the chop/nod/dither pattern placing
fewer HAWC+ overlapping pixels toward those sky directions. 
The circles surrounding the brighter stars indicate \Gaia \citep{Gaia} DR2 
\citep{Gaia_DR2} matches
(see Section~\ref{Gaia} below), colored by distance bin. 

The contours show that the
strongest thermal dust emission arises very near the YSO, with weaker resolved
emission along two thin east and west arms and a wider southwest arm. The shape of 
the extended contour
configuration is nearly identical to the distribution of C$^{18}$O shown in 
\citet[][their Figure~2, upper right]{Furuya14a}. This indicates the HAWC+ observations accurately
reveal the locations of both the L1082C core and the GF~9-2 YSO, as well as the extent of the former.

\subsection{Mimir Observations}\label{Mimir}
Observations of NIR BSP were obtained using Mimir
on the 1.8~m Perkins telescope, located in northern Arizona,
in $H$ (1.6~\um) and $K$ (2.2~\um) bands during several nights in 2011, 2013, 2014, 2015,  
and 2017. A total of 41 separate observations were obtained. Each observation consisted of
images obtained through 16 half-wave plate (HWP) orientation angles for each of 
six sky-dither positions, for
a total of 96 images. Exposure times per image were one of 2.3, 10, or 15~s. Images were
evaluated for seeing and other effects, and observations using the same exposure times obtained
during the same observing run were averaged to improve SNR and detection
sensitivity. Total exposure times were 6.6~hr in $H$-band and 8.2~hr in
$K$-band. Full details regarding Mimir polarimetry observations, data reduction, and
data products are reported in \citet{Clemens12a, Clemens12b}. 

A total of 856 stars in the Mimir field-of-view (FOV) were contained in the final, 
merged $H$-band polarization stellar catalog. 
About 15\% (125) of these stars were brighter than $m_H = 15.2$~mag, exhibited polarization SNR (``PSNR" $\equiv P^\prime / \sigma_P$\footnote{Linear polarization percentages were 
debiased using $P^\prime = \sqrt{P^2 + \sigma_P^2}$, where $P$ is the uncorrected
value and $\sigma_P$ is its uncertainty.}) greater than 1.6 (i.e., $\sigma_{PA} < 18\degr$), 
and had $P^\prime$ values less than 
4.5\% (to avoid noise-driven false positives). 
Similarly, of 688 stars in the combined $K$-band polarization
catalog, for magnitudes down to $m_K = 14.3$ there were 55 stars meeting the
same PSNR criterion with $P^\prime$ less than 4\%. 
Stars fainter than these two magnitude limits had PSNR values too 
small to yield reliable individual B-field orientations, but were able to be combined to yield useful average
properties, as described in the following Section.

\begin{splitdeluxetable}{ccCcccccccccBcccccccl}		
\tabletypesize{\scriptsize}
\tablecaption{Photometry, Polarimetry, and Parallaxes for GF~9 / L1082C Field Stars\label{tab_1}}
\tablewidth{0cm}
\colnumbers
\tablehead{
&&\multicolumn{10}{c}{Mimir Values / Uncertainties}&\multicolumn{3}{c}{2MASS Values / Unc.}&\multicolumn{2}{c}{WISE Values / Unc.}&\multicolumn{2}{c}{Gaia DR2 Values / Unc.}\\
\cline{3-12} \cline{13-15} \cline{16-17} \cline{18-19}
\colhead{No.} & \colhead{RA/decl} & \colhead{$H$} & \colhead{$P^\prime_H$} & \colhead{PA$_H$} & \colhead{$Q_H$} & \colhead{$U_H$} & \colhead{$K$} & \colhead{$P^\prime_K$} & 
\colhead{PA$_K$}& \colhead{$Q_K$} & \colhead{$U_K$} &
\colhead{$J$} & \colhead{$H$} & \colhead{$K$} & \colhead{W1} & \colhead{W2} & 
\colhead{$g$} & \colhead{$\pi$} &
\colhead{Notes}\\
&\colhead{[$\degr$]} & \colhead{[mag]} & \colhead{[\%]} &
 \colhead{[$\degr$]} &  \colhead{[\%]} & \colhead{[\%]} & \colhead{[mag]} & \colhead{[\%]} & \colhead{[$\degr$]} &  \colhead{[\%]} & \colhead{[\%]} &
\colhead{[mag]} & \colhead{[mag]} &  \colhead{[mag]} &  \colhead{[mag]} &
 \colhead{[mag]} & \colhead{[mag]} & \colhead{[mas]} &
}
\startdata
056&312.71671&11.929&  2.393&143.0&  0.659& -2.303&11.680&  1.297&135.5&  0.023& -1.310&12.868&12.006&11.753&11.621&11.622&15.779& 0.366\\ [-3pt]
     & 60.33037& 0.001&  0.109&  1.3&  0.114&  0.108& 0.006&  0.185&  4.1&  0.139&  0.185& 0.024&0.026& 0.020&0.023& 0.020&0.001& 0.038&\\
154&312.75253&13.836&  1.927&124.1& -0.752& -1.886&13.554&  2.089&116.8& -1.326& -1.797&14.576&13.836&13.595&13.450&13.469&17.683& 0.420\\ [-3pt]
     & 60.28807& 0.005&  0.638&  9.5&  0.631&  0.639& 0.007&  0.788& 10.8&  0.799&  0.782& 0.036& 0.039& 0.034& 0.024& 0.027&0.001& 0.117&\\
237&312.78150&15.426&  3.908&138.4&  0.547& -4.584&15.183&  4.170&152.2&  3.102& -4.517&16.117&15.571&15.120&14.882&14.862&19.111& 0.692\\ [-3pt]
     & 60.34538& 0.017&  2.458& 18.0&  2.359&  2.459& 0.034&  3.555& 24.4&  2.872&  3.835& 0.100& 0.142& 0.140& 0.028& 0.044&0.003& 0.261&\\
320&312.81297& 9.781&  1.377&145.8&  0.509& -1.281& 9.436&  0.713&138.0&  0.075& -0.712&10.982& 9.821& 9.520& 9.333& 9.429&14.479& 0.121\\ [-3pt]
     & 60.34628& 0.000&  0.055&  1.1&  0.050&  0.055& 0.001&  0.064&  2.6&  0.055&  0.064& 0.022&0.026& 0.022& 0.022& 0.020&0.000& 0.038&\\
816&313.01130& 7.890&  3.957&123.4& -1.585& -3.705& 7.393&  0.919&107.6& -0.762& -0.540& 7.715& 7.556& 7.510& 7.309& 7.450& 8.518& 3.407\\ [-3pt]
     & 60.37427& 0.001&  0.759&  5.5&  0.858&  0.740& 0.001&  0.163&  5.1&  0.166&  0.155& 0.018&0.053& 0.021& 0.033& 0.020&0.000& 0.028&\\
\hline
\enddata
\tablecomments{This is a shortened and selective version of the full table that is available in 
electronic form, with the rows shown here chosen to span the range of $H$-band magnitudes
for stars having complete matches to $K$, \WISE, and \Gaia values.} 
\end{splitdeluxetable}

Key quantities for the 860 stars with measured $H$- or $K$-band polarization values 
are reported in Table~\ref{tab_1}. There, stellar coordinates, $H$-band and $K$-band
photometric and polarimetric values, 2MASS \citep{Skrutskie06} photometry, \WISE \citep{Wright10} W1- and W2-Band (4.5~\um, equivalent to $M$-band) photometry, and \Gaia DR2 \citep{Gaia,Gaia_DR2} $g$-band photometry and parallaxes, as well as all associated
uncertainties, where available for each star, are presented.


\section{Analysis}

Analyses included revisiting the distance to GF~9, based on \Gaia DR2 matches to
stars with NIR polarization information, and comparison of the B-field orientations revealed
by SOFIA HAWC+ FIR polarimetry with those revealed by Mimir NIR polarimetry as well as comparison
to published $I$-band BSP and \Planck TEP. Additionally,
a test for wavelength-dependent PAs was performed,
as such a dependence had been reported previously \citep{Jones03,Poidevin06} and, 
if present, could undermine the correspondence of observed PAs with plane-of-sky
B-field orientations.

\subsection{\Gaia DR2 and the Distance to GF~9}\label{Gaia}

Within the Mimir FOV, a total of 723 stars were contained in \Gaia DR2. 
These were matched to entries in the $H$-band combined polarization
catalog, using a $0.7$~arcsec radius window, resulting in 610 matches. Plots of $H$-band $P^\prime$ versus parallax
and $(H - M)$ stellar color\footnote{$M$-band magnitudes were taken as the 
\WISE \citep{Wright10} W2-band magnitudes.} versus parallax revealed that very few stars 
in the FOV are closer than 300~pc. 
Step-wise increases in these quantities in plots versus distance for other dark clouds arise because cloud
thicknesses are much smaller than the distances to the clouds, which helps to localize the effects of 
reddending and polarization by the dust contained in such clouds \citep[e.g., Figure~13 of][]{Santos17}.
No strong steps were present in the plots of $P^\prime$ and reddening versus \Gaia distance
for the GF~9 field stars.
Additionally, as both of those quantities depend on 
dust column density, and the available directions to the stars whose light passes through the dark clouds
being studied are 
largely stochastic, the steps seen in those other clouds are often less than dramatic and the functional
forms of the steps are unknown. This
has left interpretations of such plots, for the purpose of assigning dark cloud distances, 
rather subjective.

\begin{figure}
\plottwo{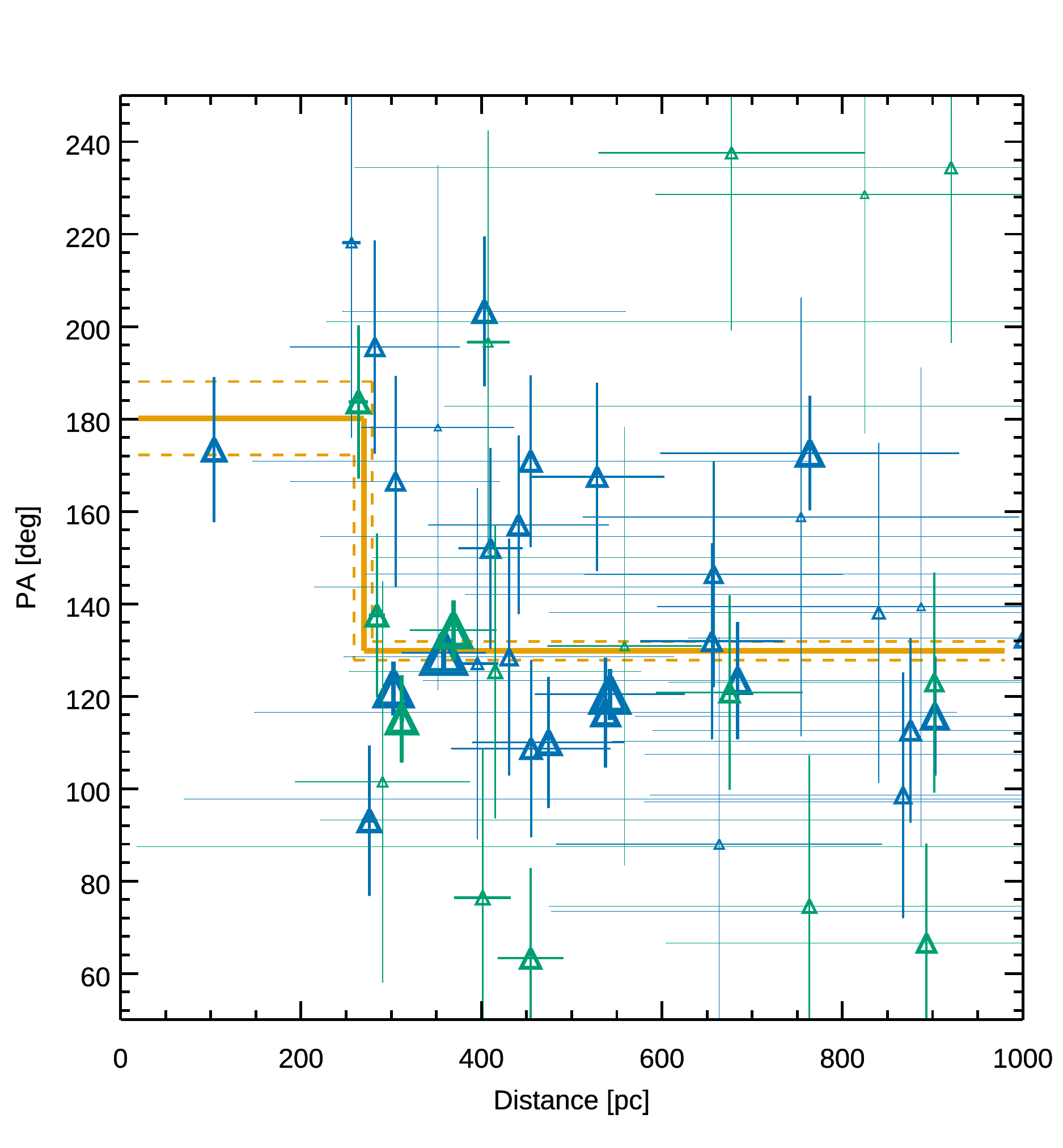}{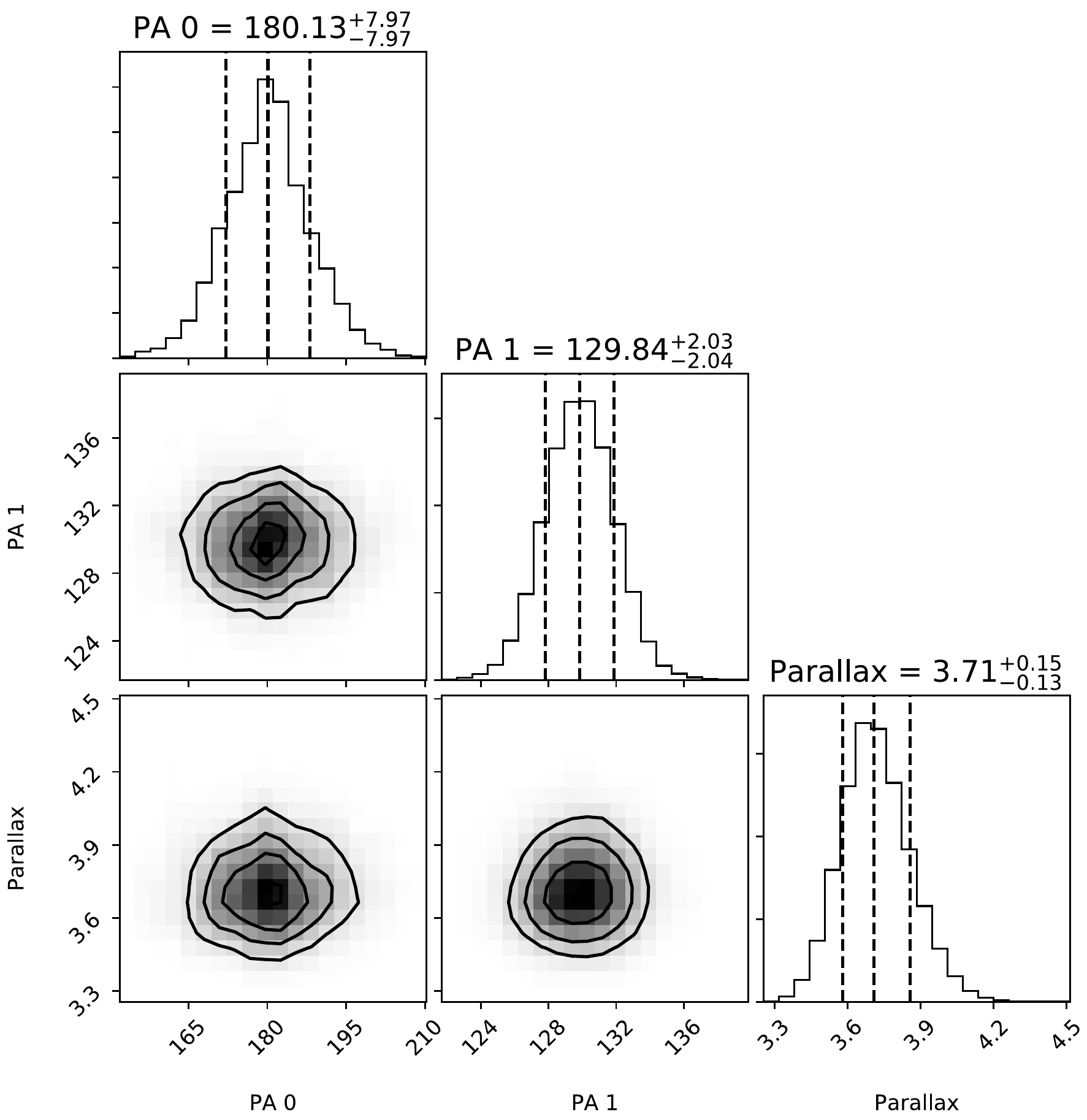}
\caption{\small
{\it (left panel)} NIR polarization PAs versus distance, based on parallaxes from \Gaia DR2, 
for the 47 stars having some reasonable
probability of being closer than 500~pc. Blue symbols and error bars represent $H$-band values and
green bars and symbols represent $K$-band values. Symbol size and thickness and error bar
thickness relate to SNR, with thicker lines indicating higher SNR. Orange lines indicate the two MCMC 
fitted PA values and the transition parallax, described in the text. Dotted orange lines identify 1~$\sigma$ certainty bands.
{\it (right panel)} MCMC corner plot showing two-dimensional likelihood distributions and marginalized
one-dimensional histograms for the two fitted PAs and the transition parallax value. The transition parallax marks the onset of the uniform PA of the B-field associated with GF~9 at $270 \pm 10$~pc. 
\label{fig_distance}}
\end{figure}

Based on the uniformity of the B-field orientations in Section~\ref{B-field} below,
a similar step change in PA might be equally definitive and simpler to 
model. 
Fifty-one stars were 
found to have values of parallax plus uncertainty in parallax greater than 2.0~mas. 
These stars were selected as having some reasonable likelihood of being located closer than 500~pc,
a value greater than the maximum of the previous distance estimates for GF~9.
In the left panel of Figure~\ref{fig_distance}, the NIR PAs measured for $H$ (in blue)
and $K$ (in green) are plotted versus \Gaia DR2 parallaxes for those stars.

A step-wise change in PA was sought using a Metropolis-Hastings Markov Chain Monte Carlo 
(MCMC) approach. The model
consisted of two uniform PA values plus a transition parallax value. Priors were uniform
in PA from 60\degr\  to 240\degr\  and uniform in parallax between 1 and 100~mas. 
Seventeen of the stars
had measured PA values with uncertainties in both $H$ and $K$ under 60\degr, so those 
PA values were combined for each of those stars, using
inverse variance weighting, to avoid double-counting the stars.
Thirty-four stars only had either $H$- or $K$-band PA uncertainties meeting 
that criterion, and so did not need to have their PA values averaged over wavelength. 
A couple of the brightest stars tended to 
dominate the initial MCMC results, so an additional 4\degr\ was added in quadrature to
all PA uncertainties to reduce that effect and permit more stars to contribute to the 
combined fit. Uncertainties in both the parallax and PA 
quantities were incorporated via gaussian likelihoods, with the parallax likelihood split 
between the two model PA values at the transition parallax. As noted in Table~\ref{tab_1},
four stars were rejected from inclusion in this fitting due to non-physical combinations of
low polarizations with large distances, oddly blue $(g - H)$ colors\footnote{The mean $g$-band
magnitudes were taken from \Gaia DR2, and are listed in Table~\ref{tab_1}.}, or strongly 
non-Serkowski \citep{Serkowski75, Wilking80} ratios
of $P_H / P_K$ (which would imply polarization processes which were not based on 
normal interstellar medium dust dichroism). Whether these represent unresolved binaries or hot stars with intrinsic
polarizations is beyond the scope of this work. 

After 500,000 steps, the MCMC routine established the stable findings shown
as the corner plot in the right panel of Figure~\ref{fig_distance}. It favors an initial
PA of $180 \pm8\degr$ that transitions to a final PA of $130\pm2\degr$
at a parallax of $3.71 \pm 0.14$~mas. None of the quantities exhibit
significant covariance, as indicated by the round two-dimensional gaussian-looking
distributions shown in Figure~\ref{fig_distance}B.  The values are indicated as the orange line in the
left panel of the plot (with uncertainty ranges indicated by the dotted orange lines).

The abrupt PA change at $270 \pm 10$~pc was adopted as 
the distance to the GF~9 cloud, the L1082C core, and the GF~9-2 YSO. Post facto weak
changes in the plots of $(H - M)$ with parallax and $P^\prime_H$ with parallax were noticed
at about the same parallax of 3.7. This correspondence is good evidence that the
distance to the cloud has been revealed via the PA step.

Of the 610 $H$-band matching \Gaia DR2 stars, only four with parallax SNR~$> 2$ have
parallaxes greater than 3.71~mas, while 253 (98\%) are more distant.
Hence the NIR BSP stars, which have a high correspondence to \Gaia DR2 stars, 
are located behind the GF~9 cloud and are ideal probes of the B-field threading the
dust in the cloud and dense core. Additionally, NIR color, $P^\prime_H$, and PA all show
an absence of step changes for distances beyond 270~pc. That is, the layer dust (and
embedded B-field) associated with GF~9 appears to be the only dust layer along
these lines of sight (L$\sim 97$\degr\ and B$\sim 10$\degr).

\begin{deluxetable}{cccccl}
\tabletypesize{\small}
\colnumbers
\tablecaption{Convolved HAWC+ 216~$\mu$m Polarization Properties\label{tab_2}}
\tablewidth{0pt}
\tablehead{
\colhead{Aperture} & \colhead{RA} & \colhead{decl.} & \colhead{$P^\prime$\tablenotemark{a}} & \colhead{BPA}&\colhead{Notes} \\
&\colhead{[$\degr$]} & \colhead{[$\degr$]} & \colhead{[\%]} &
 \colhead{[$\degr$]} 
}
\startdata
1	&	312.7908 &	60.3332 &	15 (12)	& 163 (23) & low PSNR \\
2	&	312.8174 &	60.3269 &	0.0 (6.5) & ...	 \\
3	&	312.8175 &	60.2869 &	6.2 (3.3) & 142 (15) \\
4	&	312.8329 &	60.2976 &	3.9 (2.4) & 143 (17) \\
5	&	312.8396 &	60.3215 &	2.4 (3.6)	& 121 (44) & low PSNR \\
6	&	312.8407 &	60.2863 &	3.9 (2.5) & 149 (19) \\
7	&	312.8442 &	60.3105 &	1.3 (1.8) & 124 (40) & low PSNR \\
8	&	312.8557 &	60.3009 &	0.0 (2.1) & ... \\
9	&	312.8614 &	60.2881 &	0.0 (3.5) & ... \\
10	&	312.8629 &	60.3199 &	2.6 (1.7) & 129 (18) \\
11	&	312.8697 &	60.3092 &	1.9 (1.1) & 142 (17) & YSO position\\
12	&	312.8794 &	60.2969 &	0.0 (3.5) & ... \\
13	&	312.8861 &	60.3225 &	5.3 (2.9) & 195 (15) \\
14	&	312.8913 &	60.3069 &	0.0 (3.6) & ... \\
15	&	312.9062 &	60.3153 &	0.0 (3.6) & ... \\
16	&	312.9299 &	60.3161 &	0.0 (4.1) & ... \\
17	&	312.9562 &	60.3166 &	0.0 (4.6) & ... \\
\enddata
\tablenotetext{a}{Values are followed by uncertainties in parentheses.}
\end{deluxetable}

\subsection{Convolved SOFIA Observations}\label{Obs_obs}

The extended 216~\um\ thermal dust emission from the L1082C dense core 
beyond the centrally-bright GF~9-2 YSO is very weak, with the resolved 
arm intensities shown in Figure~\ref{fig_SOFIA_Gaia}
being only $\sim$40~MJy~sr$^{-1}$ above nearby background regions. 
Smoothing to coarser 
angular resolution than the HAWC+ pixel size was necessary to boost SNR and enable polarization detections, similar
to that applied to SCUPOL \citep{SCUPOL} observations reprocessing as described in \citet{Clemens16}. 
Smoothing was 
performed on the Stokes $UI$ and $QI$ maps, weighted both by the variances in their pixel uncertainty
values and by a gaussian kernel of FWHM equal to four HAWC+ pixels. Instead of using a regular
grid of central positions for the gaussian smoothing, the intensity image 
(Figure~\ref{fig_SOFIA_Gaia}) was used as a guide for discrete synthetic-aperture 
placements, starting at
the YSO position and working along the arms seen in FIR emission. The gaussian kernel
sizes and center placements of the apertures were varied to find the maximum number of independent
apertures that exhibited PSNR~$>$~1.6. The resulting aperture-averaged Stokes $UI$, $QI$ , and
$I$ values were used to compute the polarization $P$, its uncertainty $\sigma_P$, the
corrected $P^\prime$, position angles PA and BPA\footnote{TEP returns polarization PAs that correspond to the
maximum emission orientations of aligned dust grains, which are perpendicular to the 
local B-field orientations and to the PA of BSP stars. Here, the {\underbar B}-field orientation 
{\underbar P}osition {\underbar A}ngle (BPA) is
used to signify that the TEP PAs have been rotated by 90\degr\ to become BPAs.}, 
and the uncertainty in position angle.

The  four-pixel FWHM case returned six synthetic apertures that met the PSNR criterion and
eleven that did not. The convolved HAWC+ polarization properties for these 17 apertures 
are listed in Table~\ref{tab_2} and displayed in Figure~\ref{fig_polarimetry}. Interestingly, five of the six detected apertures show 
inferred B-field orientations 
that are very close to those found for NIR BSP stars probing
nearby directions.

\subsection{B-Field Orientations}\label{B-field}
\begin{figure}
\includegraphics[trim=-0.75in 0.33in 0 0in, clip, angle=0,scale=0.75]{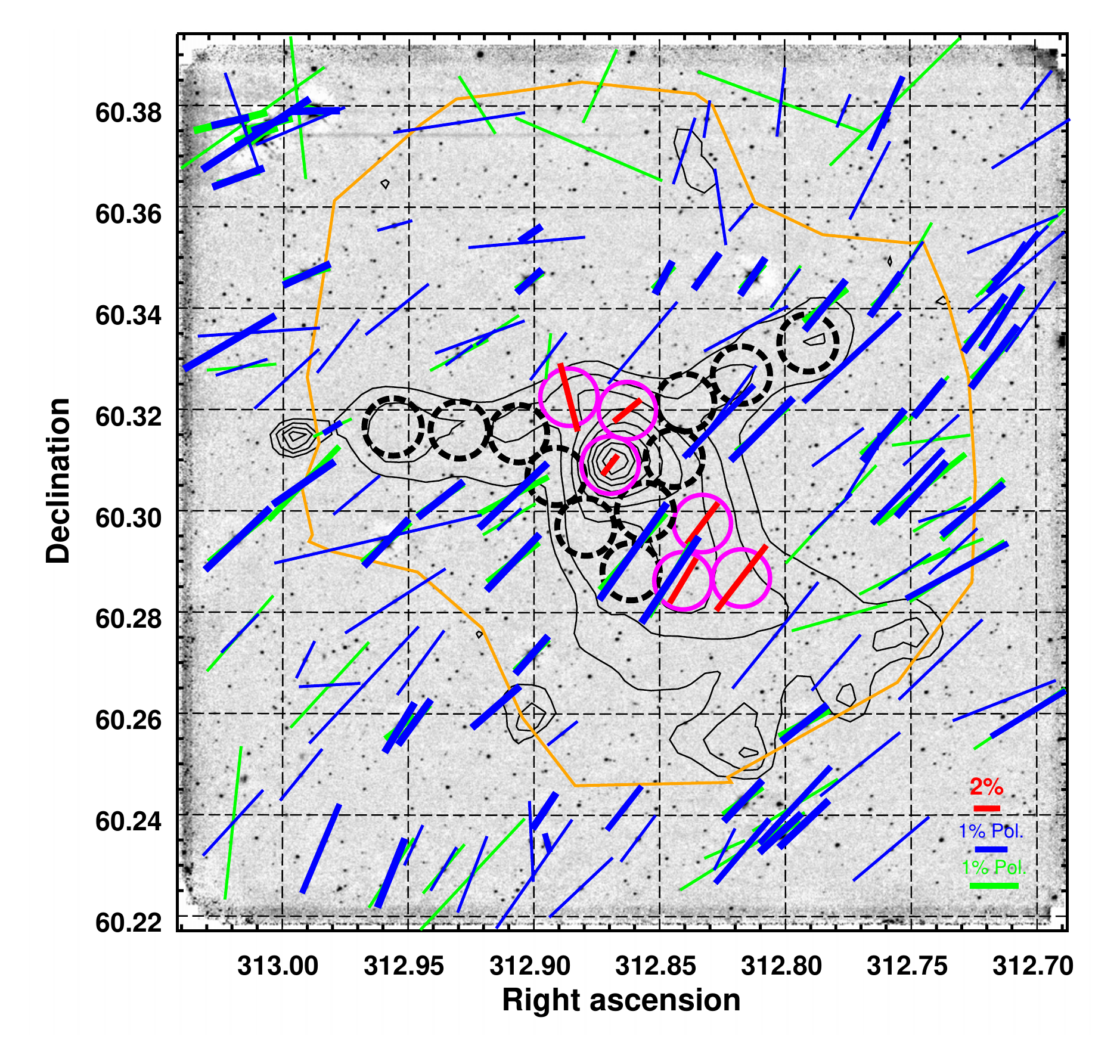}
\caption{\small
Mimir $K$-band image (from Figure~\ref{fig_SOFIA_Gaia}), with overlaid thin black contours showing 
HAWC+ 216~\um\ total intensity and orange polygon showing HAWC+ survey extent.
Blue vectors display $H$-band polarization percentages and PAs for background stars
having PSNR greater than 1.6. Green vectors show similar $K$-band values for similarly
selected stars. Thicker vectors represent greater PSNR values. 
Magenta circles with red vector inlays identify the locations and sizes
of the synthetic apertures within which significant 216~\um\ polarization was detected,
and for which the red vectors encode percentage polarizations and B-field PAs. 
The aperture closest to the 216~\um\ intensity peak is centered on the YSO coordinates.
Apertures for which significant polarization was not detected are shown as the
dashed black circles. Reference percentage polarization lengths are shown in the 
lower right corner, color coded to the the HAWC+ $E$-band, Mimir $H$-, and Mimir
$K$-band, respectively.
\label{fig_polarimetry}}
\end{figure}

There is a remarkable uniformity and agreement of the plane-of-sky B-field 
orientation BPAs seen in Figure~\ref{fig_polarimetry}, as revealed by
SOFIA HAWC+ and by Mimir polarimetry. The overall
BPA is about 135\degr, which appears to be parallel to the 216~\um\ intensity-traced
northwest arm, but is mostly 
perpendicular to the eastern arm and to the southwestern arm. 

A rough estimate of the mean column densities for each arm was developed by examining
the $(H - M)$  reddening to stars located behind the two arms.
For six stars in or near the northeastern arm, their mean $(H - M)$ is 0.66~mag, though
significantly increasingly red for the stars nearest the dense core. 
The two stars in the southeastern arm that are closest to the three SOFIA 
position detections exhibit a mean $(H - M)$ of 1.0~mag.  With typical reddening to
molecular hydrogen column density conversions, and correcting for the intrinsic
colors of the stars, the northeastern arm has a column density in the range
$3 - 4 \times 10^{21} H_2$~cm$^{-2}$ and the southeastern arm is higher, at
about $7 - 8 \times 10^{21} H_2$~cm$^{-2}$. Thus the BPA being parallel to the
elongation of the less dense northeastern arm and perpendicular to the more
dense southeastern arm follows trends seen for filamentary clouds
elsewhere \citep{Myers09, Li13, Planck_PA} for these column densities.

The mean orientation of the BPA is 
generally perpendicular to much of the large-scale dark arc of the GF~9 cloud seen in 
Figure~\ref{fig_overview}. The North Galactic Pole
direction is at PA~308.4\degr, which makes the revealed B-field for the L1082C dense core in GF~9
nearly perpendicular to the Galactic disk.

\begin{figure}
\includegraphics[trim=0.0in 0.2in 0 0in, clip, angle=0,scale=0.65]{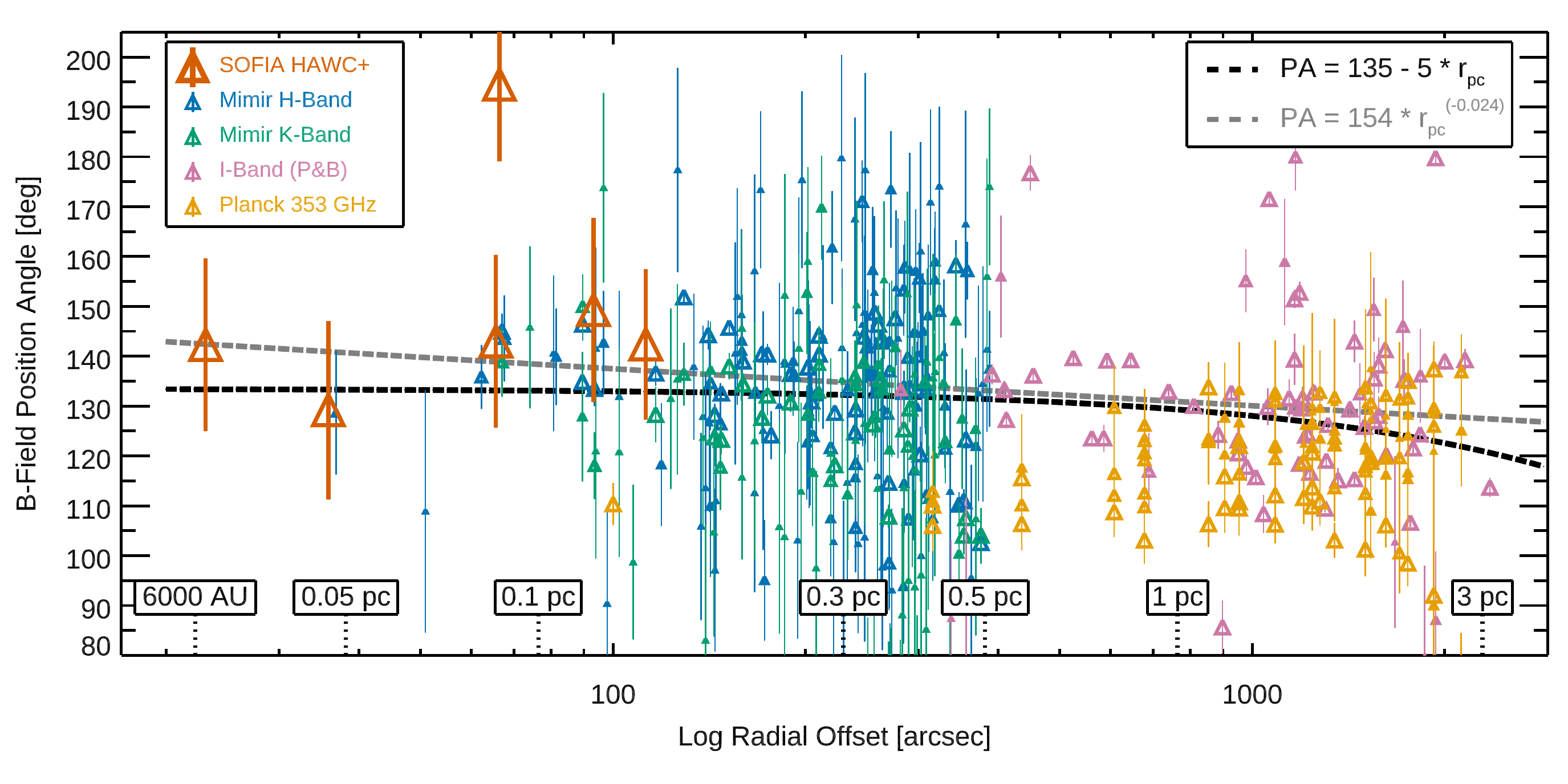}
\caption{\small
Magnetic field plane-of-sky orientation angle BPA versus the log of sky-projected radial offset
measured from the GF~9-2 YSO location, in arcsec, for SOFIA HAWC+ (red triangles and error bars), Mimir $H$ (blue) and $K$ (green), $I$-band (magenta, from \citet{Poidevin06}), and \Planck 353~GHz (850~\um, orange).
Symbol size and error bar thickness encode PSNR values, except for HAWC+ data, which are
enlarged and thickened for emphasis. Projected offset physical distances are indicated as dashed lines
connected to boxed values along the bottom axis. Central dashed black line represents a linear
fit of BPA versus offset, and the dashed gray line a power-law fit, as noted in the text, showing the 
nearly uniform B-field orientation
from 6000~AU out to more than 3~pc.
\label{fig_PA_vs_Offset}}
\end{figure}

The degrees of B-field uniformity and agreement were probed by comparing the BPAs versus offset from the YSO position for these FIR and NIR
data sets. In addition, the $0.77$~\um\ ($I$-band) BSP of \citet{Poidevin06}
and the \Planck~353~GHz (850~\um) TEP, both of which span much larger fields-of-view, were
included in the comparison. All these data points were selected subject to a position angle
uncertainty criterion of less than 25\degr. After selection, two different weighted fits 
of BPA versus offset
were performed. These data, and the fits, are shown as Figure~\ref{fig_PA_vs_Offset}.

In that Figure, the TEP position offsets have been added in quadrature with half
of their FWHM beamsizes, to indicate their effective beam-averaged offsets from the YSO
position. Symbol color identifies the data set, as listed in the legend in upper left. 
Some key physical offset values are marked and listed just above the lower axis. 

Fits of two different forms for BPA versus offset were performed, one was linear in BPA and offset,
the other was a power-law relation.
In the Figure, the black-dashed line represents the linear fit of the full set of BPAs versus offset, weighted by
the variances of their uncertainties, as:
\begin{equation}
{\rm BPA}_{deg} =  (134.9 \pm 0.3) -  (4.6 \pm 0.2) \times r_{pc},
\end{equation}
where BPA$_{deg}$ is the B-field equatorial orientation angle measured in degrees and $r_{pc}$ is the radial 
offset measured in parsecs.
The gray-dashed line represents the power-law fit:
\begin{equation}
{\rm BPA}_{deg} = (153.6 \pm 0.6) r_{pc}^{(-0.024 \pm 0.003)}.
\end{equation}
The weak slopes for both of these fits, less than 5\degr\ per pc for the linear fit, confirms 
the high degree of uniformity of the mean
BPA across two orders of magnitude of physical offset. 

There are some deviations away from the uniform mean B-field. Some are seen in Figure~\ref{fig_PA_vs_Offset}
as the scatter off the mean, by up to about 25\degr\ for the farthest offset
Mimir and $I$-band
data. In Figure~\ref{fig_polarimetry}, the blue
$H$-band Mimir vectors show small PA undulations along the mean field
when traversing from the lower left to upper right. Additionally, one SOFIA HAWC+ position
shows an extreme departure from all other SOFIA BPAs and from the local Mimir
BPAs. 

This last BPA departure could be an early indication of changes in the local B-field
direction driven by dense core collapse onto the envelope surrounding
a YSO disk. Alternatively, it could be due to the disruption caused by the weak outflow from 
the YSO, which itself shows some elongation to the northeast \citep[PA$\sim 45$\degr;][]{Furuya14b}, 
the same as the PA of the line connecting
the YSO to that SOFIA position. Faint emission is seen in \Spitzer IRAC images to $\sim$35~arcsec
southwest of the YSO (along PA$\sim225$\degr). 
If this emission represents the extent of the other outflow lobe,
the lack of SOFIA HAWC+ detection in this area could be explained by disrupted or tangled B-field lines.

\subsection{PA Rotation with Wavelength}
The examination for rotation of BPA
with wavelength was performed in two ways. The first was a non-quantitative examination
of the data presented in Figure~\ref{fig_PA_vs_Offset}. In that figure, there is a slight
tendency for the HAWC+ points to exhibit greater BPA values while the \Planck values
are lesser. However, there is no strong progression of BPA with wavelength that stands
out at any radial offset. The Figure also shows that because the different 
polarization probes each span limited radial offset zones, detailed quantitative
comparisons that use the full data set for each wavelength will introduce a spatial bias,
since there is no single zone sampled by all wavebands.

\begin{figure}
\includegraphics[trim=-2.5in 0.2in 0 0in, clip, angle=0,scale=0.65]{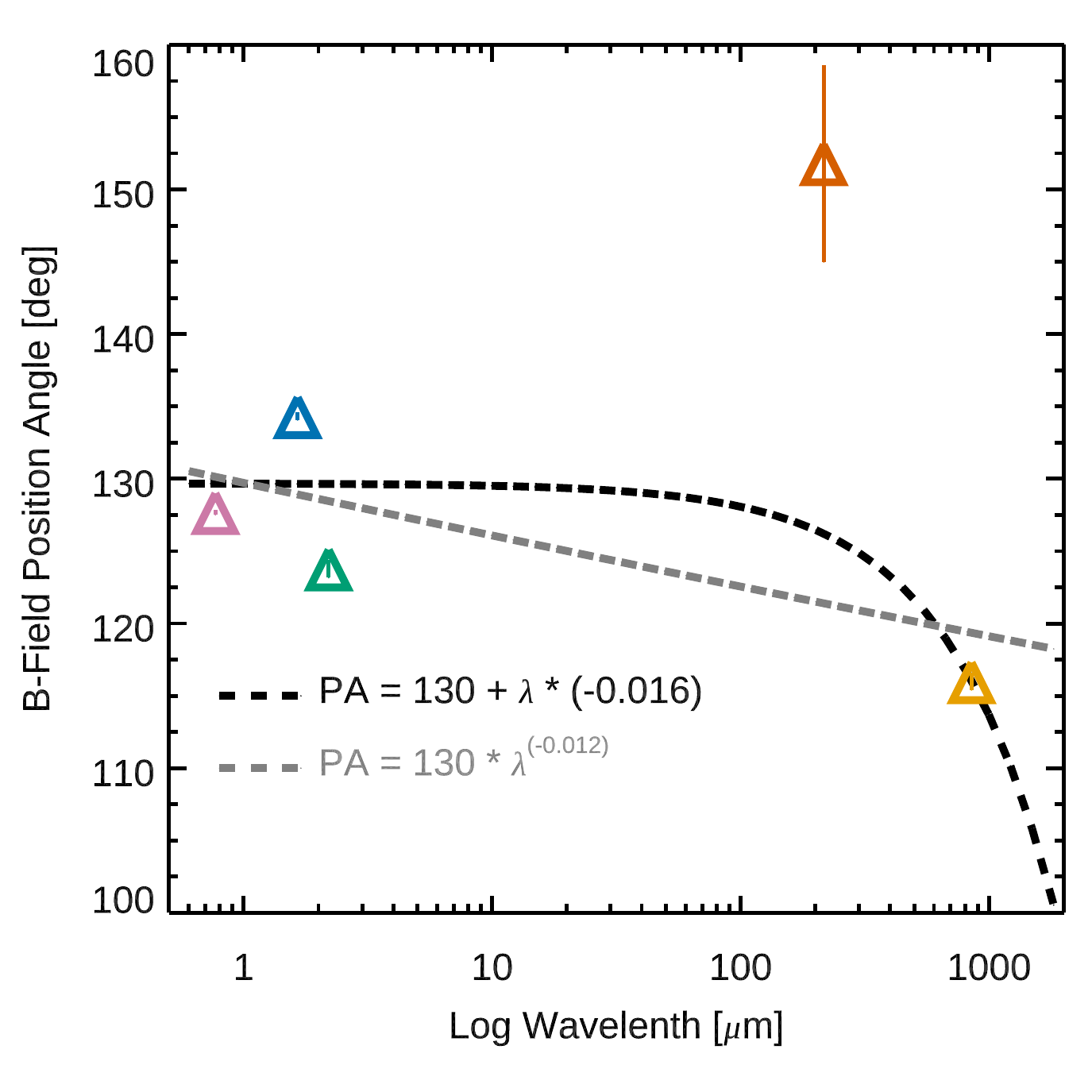}
\caption{\small
B-field position angles versus log of wavelength for $I$-band (magenta triangle), $H$-band 
(blue), $K$-band (green), HAWC+ 216~\um\ (red), and \Planck 353~GHz (orange) data set
weighted averages. 
Uncertainties are displayed as vertical colored lines, though only the 216~\um\ line extends
outside its plot symbol.
Two types of notional least squares fits are indicated. The dashed black
curve is a linear fit of BPA versus wavelength and the dotted gray curve is a power-law
fit. Both fits indicate only very weak departures with wavelength from the mean BPA.
\label{fig_PA_vs_wave}}
\end{figure}

Nevertheless, in the second examination method, weighted-average BPA values were 
formed for the data set at each wavelength, using a common criterion that each data set
utilized only elements for which the BPA 
uncertainties were less than 25\degr.
This resulted in sample sizes of 73 stars for $I$-band, 183 stars for $H$-band,
115 stars for $K$-band, six positions for HAWC+ 216~\um, and 115 positions for
\Planck~353~GHz out to about 0.5\degr\ offset from the YSO. The mean BPA values, and 
propagated uncertainties, are shown in Figure~\ref{fig_PA_vs_wave} as colored triangles
and vertical error bars. All six SOFIA BPAs were used, including the one which significantly
departs from the other five. If only the five are used, the SOFIA data point moves down
on the plot by 10\degr. Also, some of the \Planck data correspond to offsets 
that are quite far from 
the YSO position, so even a weak change in BPA with location will affect the utility of 
inclusion of the \Planck data with the SOFIA and stellar data for the purpose of finding
a wavelength dependence.

Least-squares fits to the run of BPA with wavelength were performed; one 
was linear with wavelength and one was a power-law of wavelength. The fit functions and
parameters returned were:
\begin{equation} 
{\rm BPA}_{deg} = (129.7 \pm 0.2) - (16.0 \pm 0.7)\times 10^{-3} \times \lambda_{\mu m}, 
\end{equation}
{\rm and}
\begin{equation}
{\rm BPA}_{deg} = (129.72 \pm 0.10) \times \lambda_{\mu m}^{(-0.012 \pm 0.002)},
\end{equation}
where BPA$_{deg}$ is the B-field orientation angle in degrees and $\lambda_{\mu m}$ is the
wavelength measured in microns.
While these fits do, formally, report BPA rotations with wavelength, the rotation amounts 
are quite small, amounting to less than $15$\degr\ over the entire wavelength 
range. In contrast to the BPA rotation with wavelength (and/or position) identified by \citet{Jones03} and \citet{Poidevin06} for the L1082C core in GF~9, Figures~\ref{fig_polarimetry} and \ref{fig_PA_vs_wave} show negligible BPA rotation with {\it either} wavelength or position. 

\section{Discussion}

The \Gaia parallax plus polarization BPA-based distance determination of 270~pc leads to only a minor
revision of the inferred YSO luminosity, to about 0.5~$L_\odot$, thus affirming that this YSO is 
one of the lowest
luminosity Class~0 YSOs. This new distance also points to a minimal or absent foreground
extincting  dust layer and places nearly all of the stars measured for NIR polarization
behind the cloud, confirming their utility for revealing the B-field of the L1082C core region. Further,
the comparisons of \Gaia distances versus reddening and versus polarization indicated an absence of any significant
dust layers more distant than GF~9 in this direction, helping isolate the dense core for B-field characterizations.

Background starlight polarization has been used to trace B-fields in the envelopes
of other dark clouds \citep[e.g.,][]{Goodman90}. This technique was
extended to the NIR \citep{Goodman92, Creese95} 
to try to probe the more extincted regions where star formation 
might occur. Comparisons of the NIR BSP values
with those obtained at optical wavelengths \citep{Goodman95, Creese95} 
led to the conclusion that B-fields in cloud interiors are not being 
sensed by BSP. Explanations favored either the B-field being located only in 
a skin around dark clouds \citep{Goodman95} or a 
change in dust grain properties between the cloud envelopes and their interiors 
\citep{Lazarian97}, which could cause reduced grain 
alignment efficiency with the local B-fields \citep{Creese95}. 

While not analyzed in detail here, the L1082C core {\it does} seem to be threaded by a 
B-field that is present in more than just the low-density periphery of the core. 
This conclusion is supported by the increased 
NIR polarization $P^\prime$ for stars probing close to the core 
(Figure~\ref{fig_polarimetry}), the close correspondence of the BPAs for both the
NIR BSP and the FIR TEP (the latter of which are insensitive to the low density cloud
periphery), and in the decreased FIR $P^\prime$ nearest
the YSO, likely due to field line tangling and associated depolarization for B-field configuration
changes closer to the YSO than the 6000~AU probed here \citep[e.g.,][]{Kataoka12, Lee17}.

Indeed, the long, threaded-filamentary structure of the GF~9 dark cloud,
which spans some 8~pc in length, varies in projected width from 0.1 to 0.4~pc,
and hosts four or more embedded low-mass star formation zones 
within the multiple dense 
cores inside the cloud \citep{Furuya08}, is well-matched to the collapse conditions simulated by multiple
recent studies \citep[e.g.,][]{Kataoka12,Seifried15,Chen16,Hull17,Lee17}.

Here, the main finding of a predominantly uniform plane-of-sky projected B-field from very
near the YSO ($\sim6000$~AU) to the outer, diffuse 3~pc zone examined in the 
\Planck polarimetry is at odds with many models employing B-fields that are weak to moderate,
in comparison to local turbulent energies. That is, the B-field structure of the GF~9/L1082C region
is not super- or trans-Alv{\'e}nic ($\mathcal{M}_A > 1$).
In those models, the B-fields are 
dragged along with coherent or turbulent gas dynamic motions
\citep{Seifried15,Hull17} to produce non-uniform polarization patterns. 
The \citet[][see their Figure~2]{Hull17} work includes simulations of resulting polarizations
for size scales from 5~pc down to about 300~AU, fully encompassing the range of
GF~9 size scales probed here. Of their simulations, the only one that
bears any relation to the uniform B-field seen for GF~9 is the 120~$\mu$G, 
strong-field (sub-Alv{\'e}nic, $\mathcal{M}_A < 1$) case.
All others cases predict polarization PA departures from uniformity that would have been detected in 
the SOFIA or 
Mimir observations of GF~9.

The isolated, low-mass star formation taking place in GF~9 and the near-perpendicular
orientation of the B-field to the elongation direction of the main filament
argue for a strong field condition ($\mathcal{M}_A < 1$) dominating the evolution of GF~9.

The lack of wavelength dependence in the projected BPAs from $I$-band to
850~\um\ removes any remaining concern that the PAs might not be tracing the bona fide
B-field orientations. The earlier claims of such a dependence rested on the \ISO polarimetry
findings, which were calibrated using assumptions judged to be questionable post facto, 
but which now may be tied to \Planck PA values to recover useful
BPAs in the future (an effort beyond the scope of this paper), and some unknown error 
in the earlier NIR polarimetry. No rapid variation in BPA with wavelength was found using
Mimir, \Planck, and SOFIA polarimetry.


\section{Summary}

SOFIA HAWC+ 216~\um\ polarimetry, which sampled the thermal dust emission from 
the L1082C dense core and YSO environs of GF~9-2, was combined with 
Mimir NIR polarimetry of stars located behind the GF~9 cloud, as well as with previously
published optical and \Planck polarimetry.  These combined data were used to develop a 
comprehensive characterization
of the plane-of-sky orientation of the B-field that threads the cloud, the dense core,
and the YSO region. 

\Gaia DR2 stellar distances were examined to refine the distance determination
to GF~9 and the YSO. A new method that combines polarization PAs,
\Gaia parallaxes, and a Bayesian MCMC approach applied to a step-wise model
revealed a single
PA transition at $270 \pm 10$~pc, which was adopted as the distance to the
cloud core. The \Gaia distances, combined with $(H - M)$ reddenings and NIR 
polarizations, revealed that GF~9 contains the sole extincting layer along this direction, and
that nearly all of the NIR polarization stars (98\% of the \Gaia stars) are located 
beyond GF~9, and so satisfy a key criterion for background starlight 
probes of B-fields via dichroic polarimetry.

The B-field traced by the SOFIA HAWC+ FIR emission polarimetry and the Mimir NIR background starlight
polarimetry is remarkably uniform, changing very
little from the outer, diffuse region of the cloud, some 3~pc from the YSO, to the
smallest size scales, $\sim$6000~AU from the YSO. 
\deleted{Even the recently found small outflow, of size 1,000~AU, appears to be perpendicular to the mean
B-field, implying that the uniform B-field orientation condition also may be present down to 
this smaller size scale.}
Comparing the B-field 
configuration to recent simulations favors a strong-field ($\mathcal{M}_A < 1$) condition during the star
formation process that led to the GF~9-2 YSO.

No strong change was found in the B-field orientations with wavelength, across the full range
from 0.77~\um\ to 850~\um, counter to earlier reports, thus removing the concern
that polarization PAs might not reveal B-field orientations in this setting.

The combination of SOFIA HAWC+ FIR polarimetry, which probed the
densest cloud core regions, with Mimir NIR polarimetry, which probed dusty
material around the core and out to the diffuse cloud edge, was an effective multi-scale
tool for tracing B-field properties in the GF~9 cloud, the L1082C dense core, and 
the environs of the GF~9-2 YSO as they undergo the early stages of 
star formation.

\acknowledgments

Darren Dowell, the entire SOFIA HAWC+ team, the SOFIA flight and ground crews, 
and the USRA SOFIA Project teams developed the SOFIA observatory, the HAWC+
instrument, performed the airborne observations, processed and calibrated the
GF~9-2 data, and delivered science-ready data products. AEB and J. Montgomery thank the guest observer flight support team for aiding their
participation in the SOFIA flights for this project.

Based in part on observations made with the NASA/DLR Stratospheric Observatory for Infrared Astronomy (SOFIA). SOFIA is jointly operated by the Universities Space Research Association, Inc. (USRA), under NASA contract NNA17BF53C, and the Deutsches SOFIA Institut (DSI) under DLR contract 50 OK 0901 to the University of Stuttgart. Financial support for this work was provided by NASA through award SOF~04\_0026 issued by USRA.

This work has made use of data from the European Space Agency (ESA) mission
{\it Gaia} (\url{https://www.cosmos.esa.int/gaia}), processed by the {\it Gaia}
Data Processing and Analysis Consortium (DPAC,
\url{https://www.cosmos.esa.int/web/gaia/dpac/consortium}). Funding for the DPAC
has been provided by national institutions, in particular the institutions
participating in the {\it Gaia} Multilateral Agreement.

This publication makes use of data products from the Two Micron All Sky Survey, 
which is a joint project of the University of Massachusetts and the Infrared 
Processing and Analysis Center/California Institute of Technology, funded by 
NASA and NSF. 

This publication makes use of data products from the Wide-field Infrared Survey Explorer, which is a joint project of the University of California, Los Angeles, and the Jet Propulsion Laboratory/California Institute of Technology, funded by the NASA

Based on observations obtained with \Planck (http://www.esa.int/Planck), an ESA science mission with instruments and contributions directly funded by ESA Member States, NASA, and Canada.

In addition to the authors, Mimir observations were conducted 
in part by A.~Pinnick, J.~Moreau, R.~Marchwinski, M.~Bartlett, and C.~Trombley. 
Participation by GS was supported in part by an NSF REU Site grant, number AST13-5920,
to Boston University (M. Opher, PI).
This research was conducted in part 
using the Mimir instrument, jointly developed at Boston University and Lowell 
Observatory and supported by NASA, NSF, and the W.M. Keck Foundation.
This effort has been made possible by grants 
AST 06-07500, AST 09-07790, AST 14-12269 from NSF/MPS,
USRA SOF\_4-0026, and NASA NNX15AE51G to Boston University and by grants of 
Perkins telescope observing time from the Boston University 
-- Lowell Observatory partnership.

\facility{Perkins, SOFIA}

\end{document}